# Title

*Complete Trip: A Linked Multimodal Human Mobility Dataset*

## Authors

Ruohan Li[1*], Weiyu Luo[1*], Xin Wu[1], Chenfeng Xiong[1†]

**Affiliations**

1. Department of Civil & Environmental Engineering
Villanova University, Villanova, PA 19085
*These authors contributed equally to this work
†Corresponding Author

Corresponding author(s): Chenfeng Xiong (chenfeng.xiong@villanova.edu)

## Abstract

Human mobility data have become fundamental to research across transportation, public health, urban science, and disaster resilience. However, existing mobility datasets typically capture only isolated aspects of travel behavior and rarely provide linked multimodal journeys together with network-level route representations and population-level inference. Here we present Complete Trip, a mobility dataset that reconstructs linked multimodal travel behavior from passively collected smartphone location-based services (LBS) data. The first released implementation covers six counties in Utah throughout 2020 and represents journeys across car, bus, rail, and active transportation through a four-stage workflow consisting of trip identification, mode imputation, route reconstruction, and trip linking. Complete Trip preserves journey-level relationships by linking sequential travel segments where multiple segments belong to the same travel episode, provides network-based route representations on digital transportation networks, and supports population-level analyses through statistically calibrated expansion weights. By providing a representation of linked multimodal human mobility, Complete Trip enables reproducible research across transportation, public health, urban science, disaster resilience, and related fields.

## Background & Summary

Understanding how people move, when, where, by what means, and through which routes, has become a foundational requirement across a wide range of scientific and policy domains. In transportation planning, this knowledge underlies multimodal accessibility assessment, transit network design, and travel demand modeling (Sinclair and Schaefer, 2019; Zhou et al., 2021; Hadi et al., 2022). In public health, the COVID-19 pandemic highlighted the critical role of individual mobility in modeling disease transmission and evaluating the effectiveness of mobility-based interventions, while similar data continue to support broader epidemiological research beyond the pandemic (Xiong et al., 2020; Hu et al., 2021; Liu et al., 2025). In urban science, individual travel and activity sequences reveal the interactions between the built environment and human behavior, informing land-use planning, spatial equity analysis, and city-scale simulation (Gonzalez et al., 2008; Song et al., 2010). Mobility data are also essential for disaster response and resilience planning, enabling researchers to understand how populations evacuate, reroute, and recover under disrupted transportation networks (Zhang et al., 2026). Across these domains, a common requirement emerges: the ability to capture complete travel behavior rather than fragmented observations of individual trips, travel segments, or locations.

The evolution of human mobility data has closely followed the increasing demand for understanding travel behavior. Early mobility observations relied on diverse but largely isolated data sources, including infrastructure-based sensing systems, mobile network records, household travel surveys, GPS tracking, and commercial app-based location data. Collectively, these data sources substantially advanced mobility research, but they were primarily designed for specific applications and therefore captured complementary rather than comprehensive views of travel behavior.

In response to demand for richer mobility information, numerous open mobility datasets have emerged in recent years, each advancing different aspects of travel behavior. Aggregate mobility products, such as Google Mobility Reports (Google LLC, 2020) and SafeGraph (SafeGraph, 2022), substantially expanded spatial coverage but primarily provide aggregated mobility indicators rather than linked individual journeys. Trajectory datasets, including GeoLife (Zheng et al., 2008) and YJMob100K (Yabe et al., 2024), capture detailed or large-scale movement trajectories but generally lack multimodal information, network-based route representations, and population-level inference. Survey- and panel-based datasets, such as PFLOW (Kashiyama et al., 2017) and MOBIS (Molloy et al., 2023), provide richer behavioral information, including multimodal travel and linked trip records, but remain geographically constrained or are not designed for population-representative mobility estimation. Synthetic datasets, such as Synmob (Zhu et al., 2023), facilitate methodological development but do not directly represent observed travel behavior. As summarized in **Table 1**, these datasets provide complementary capabilities, yet no publicly available dataset simultaneously combines multimodal travel, linked trips, network-based route representations, and population representation within a unified framework.

**Table 1** Comparison of representative mobility datasets across four core Complete Trip capabilities.

| Dataset | Multimodal | Linked trips | Network representation | Population representation |
|---|---|---|---|---|
| GeoLife (Zheng et al., 2008) | ○ | ○ | ○ | ○ |
| PFLOW (Kashiyama et al., 2017) | ● | ● | ○ | ● |
| Google Mobility (Google LLC,2020) | ○ | ○ | ○ | ○ |
| SafeGraph (SafeGraph, 2022) | ○ | ○ | ○ | ○ |
| Synmob (Zhu et al., 2023) | ○ | ○ | ○ | ○ |
| MOBIS (Molloy et al.,2023) | ● | ● | ○ | ○ |
| YJMob100K (Yabe et al., 2024) | ○ | ○ | ○ | ○ |
| Complete Trip | ● | ● | ● | ● |

*Notes: Filled circles indicate explicitly released capabilities. "Multimodal" denotes trip-level mode labels; "linked trips" denotes relationships among sequential trip records; "network representation" denotes routes referenced to transportation networks; and "population representation" denotes statistical expansion weights.*

We define a **Complete Trip** as a linked, multimodal representation of an individual's travel behavior that preserves the sequential relationships among travel segments, transportation modes, transfers, and network routes within a single journey. To operationalize this concept, we introduce **Complete Trip**, a mobility dataset reconstructed from passively collected smartphone location-based services (LBS) data. The first released implementation covers six counties in Utah throughout 2020 and reconstructs trip-level journeys across car, bus, rail, and active transportation, with explicit linking where multiple segments belong to the same travel episode. Complete Trip is generated through a four-stage workflow consisting of trip identification, mode imputation, route reconstruction, and trip linking, following the NextGen National Household Travel Survey (NextGen NHTS) framework developed by the U.S. Federal Highway Administration (FHWA, 2021). Rather than releasing isolated trajectories, origin-destination (OD) records, or aggregated mobility indicators, Complete Trip provides a unified representation of complete travel behavior through linked multimodal journeys, network-based route representations, and population-level inference.

Complete Trip introduces three key capabilities that distinguish it from existing mobility datasets.

- First, it represents complete travel behavior as linked multimodal journeys, preserving the sequential relationships among travel segments, transportation modes, and transfers within each journey.
- Second, it supports population-level inference through statistical expansion weights that correct for known sampling biases in location-based services data.
- Third, it represents trips directly on digital transportation networks, enabling network-level analyses beyond raw trajectory observations.

The following sections describe the generation pipeline, released data records, technical validation, and example applications of Complete Trip.

## Dataset Generation Framework

The Complete Trip dataset is produced through a four-stage generation framework, as illustrated in **Figure 1**. Beginning with raw LBS observations, the pipeline performs mobility reconstruction to identify trips, impute travel modes, reconstruct routes, and link trips. Population expansion weighting is then applied to improve correspondence with the broader traveling population. Concurrently, a privacy preservation framework removes or generalizes potentially identifying attributes prior to public release. The following sections describe each stage in turn.

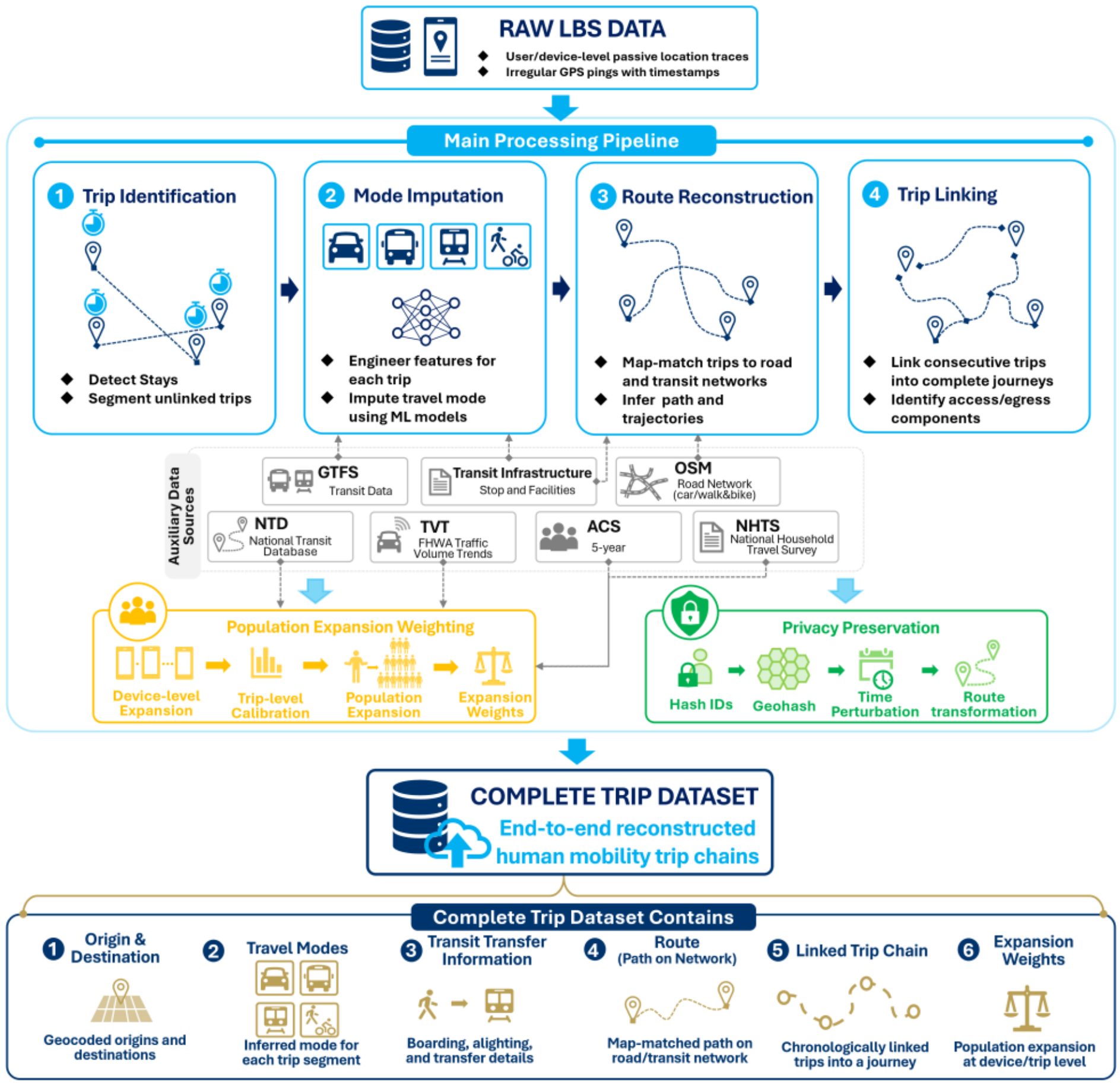

**Figure 1** Dataset Generation Framework for the Complete Trip Dataset.

### Raw LBS data

#### *Source* and observation unit

The source mobility data were provided by a commercial LBS data vendor under a data use agreement. The vendor aggregates passively collected GPS signals from mobile applications with user-granted location permissions. The fundamental observation unit is a device sighting, defined as a timestamped geographic coordinate (ping) emitted by a smartphone or GPS-enabled device during movement. Each raw sighting record contains an anonymous device identifier, a UTC timestamp, and a WGS84 latitude-longitude coordinate pair.

For the Utah study area covering six counties (Salt Lake, Davis, Weber, Utah, Wasatch, and Summit), the raw LBS dataset spans the full calendar year of 2020. **Table 2** summarizes the overall characteristics of the raw dataset, including its scale, sampling intensity, and temporal observation characteristics. As summarized in **Table 2**, the raw dataset contains more than 6.3 billion location observations from over 14.8 million anonymous devices. Despite its large scale, the raw observations are inherently sparse and event-driven, reflecting passive smartphone location reporting rather than continuous GPS tracking. These characteristics motivate the subsequent processing pipeline for trip reconstruction, mode inference, trip linking, and population expansion.

**Table 2.** Summary characteristics of the raw LBS observations.

| Attribute | Value |
|---|---|
| Total raw pings | 6,385,490,392 |
| Annual unique observed devices | 14,812,042 |
| Mean monthly active devices (MAU) | 1,828,453 |
| Mean monthly regularly active devices (RAU) | 154,134 |
| Mean monthly RAU/MAU ratio | 9.84% |
| Device-months observed on only one day | 65.46% |
| Median pings per active day among RAU devices | 60 |
| Median observed days per RAU device-month | 22 |

Note: Regularly active users (RAU) were defined as device-months observed on at least seven distinct days, with at least ten valid observations on each qualifying day (FHWA, 2021). Annual unique devices include transient and sparsely observed devices and should not be interpreted as a continuously observed population panel.

### *Sampling characteristics and potential bias*

Consistent with previous studies, passively collected LBS data do not constitute a representative sample of the underlying population. Older adults, children, low-income households, and individuals without smartphones are generally underrepresented (De Montjoye et al., 2013; Yin et al., 2015; Monreale and Pellungrini, 2023). To improve population representativeness, the Complete Trip dataset subsequently applies a population expansion weighting procedure to mitigate known demographic sampling biases. Nevertheless, weighting cannot fully compensate for populations that generate no LBS observations, such as non-smartphone users, and residual sampling bias remains an inherent limitation of passively collected smartphone mobility data.

## Mobility reconstruction pipeline

### *Trip identification*

Raw LBS observations were filtered to remove duplicate and low-quality records and were then organized into travel episodes using inferred activity anchors and the FHWA NextGen National Household Travel Survey (NextGen NHTS) trip-identification framework (FHWA, 2021). Consecutive movement observations were segmented into individual trips, producing trip-level origin, destination, departure-time, and arrival-time attributes for subsequent mode inference and route reconstruction.

### *Mode imputation*

Each identified trip was assigned to one of four travel modes: car, bus, rail, or active transportation (walk/bike). Travel mode inference was performed using a supervised Random Forest (RF) classifier trained on labeled multimodal travel data following the NextGen NHTS methodology (FHWA, 2021; Yang et al., 2022). The classifier used trip-level behavioral characteristics together with transportation-network context derived from road and transit datasets.

### *Route reconstruction*

To facilitate network-based transportation analyses, each trip was reconstructed on the corresponding transportation network according to its inferred travel mode. Rather than retaining only origin and destination locations, the reconstruction process generated the most probable network path together with route-level attributes.

For car and active transportation trips, trajectories were map-matched to the OpenStreetMap (OSM) road and pedestrian networks using Sup-HMM, an enhanced Hidden Markov Model specifically developed for sparse LBS trajectories (Xiong et al., 2018; Li et al., 2024). The reconstructed routes were represented as ordered sequences of OSM node identifiers, from which network distance and other route attributes were derived.

Transit trips were reconstructed using the General Transit Feed Specification (GTFS)-derived transit network. The resulting records included the complete transit path together with access and egress stops, transfer information, and network-based travel distance.

These mode-specific reconstruction procedures transformed reconstructed trips into network-referenced travel trajectories, providing the network representation required for subsequent complete-trip linking while enabling direct use of the released dataset in network-based transportation analyses without additional map matching.

### *Trip linking*

Following trip identification, mode imputation, and route reconstruction, each reconstructed trip initially represented a single-mode travel segment. A trip linking procedure was subsequently applied to reconstruct complete travel journeys by linking adjacent trip segments that belong to the same travel episode. The linking procedure followed the NextGen NHTS methodology (FHWA, 2021) and preserved both multimodal travel sequences and transfer relationships. Each reconstructed journey was assigned a unique linked journey identifier (`linked_trip_id`), while all constituent trip segments retained their original unique trip identifier (`trip_id`). The resulting linked-trip structure formed the final Complete Trip dataset, enabling analyses of complete travel journeys while preserving individual trip segments for multimodal, transfer, and first-/last-mile analyses.

## Population Expansion Weighting

To improve the population representativeness of passive LBS data, a two-stage weighting framework was applied to the released dataset. The framework followed the standardized methodology developed for NextGen NHTS (FHWA, 2021), while all expansion factors were estimated specifically for the Utah 2020 dataset released in this study.

### *Device-level expansion*

The first stage estimated device-level expansion factors to account for demographic and geographic sampling biases inherent in passive LBS data. Each device was assigned inferred age and household income groups. Expansion factors were then estimated through county-level iterative proportional fitting (IPF), aligning the weighted device distribution with ACS population control totals.

### *Trip-level calibration*

Device-level expansion alone did not fully correct mode-specific sampling biases inherent in passive LBS data (Wang and Chen, 2018; Wang et al., 2019). Therefore, a second-stage trip-level calibration was applied separately for driving, transit, and active transportation. Calibration factors were derived from independent national benchmark datasets, including the 2017 NHTS (National Household Travel Survey, 2025), FHWA Traffic Volume Trends reports (FHWA, 2025), the National Transit Database (U.S. Department of Transportation, 2025), and U.S. Census population estimates, enabling the weighted trips to better represent 2020 travel activity across different transportation modes. The resulting expansion factor is provided in the `trip_weight` field for eligible records.

### Privacy Preservation

The released Complete Trip dataset incorporates multiple privacy-preserving measures to minimize the risk of re-identification. All device identifiers are anonymized prior to processing, and the released data further employ spatial generalization, temporal abstraction, and route transformation to reduce the risk of disclosure of individual travel behavior while preserving population-level mobility patterns.

#### *Spatial generalization*

Trip origins and destinations are generalized to Geohash-6 grid cells rather than released as precise geographic coordinates. This spatial aggregation prevents identification of specific residences, workplaces, and other frequently visited locations.

#### *Temporal abstraction*

Exact travel timestamps are not released. Trip start and end times are represented as 30-minute time intervals, while calendar dates are generalized to the day of week and month.

#### *Route transformation*

Raw GPS trajectories are not included in the released dataset. Instead, trips are represented as inferred network paths reconstructed through the mode-specific routing procedures described above.

## Data Records

The released Complete Trip dataset contains privacy-preserving multimodal trip records reconstructed from passive LBS observations across six Utah counties in 2020. Records from the same reconstructed journey share a common `linked_trip_id`, and the dataset provides inferred travel modes, network-referenced routes, transit access and egress information, and population expansion weights for eligible records. **Table 3** summarizes the overall characteristics of the released Complete Trip dataset, including its spatial and temporal coverage, data volume, multimodal composition, network representation, and population expansion capability. The 282,508 observed devices correspond numerically to approximately 10.93% of the estimated six-county resident population; this ratio should not be interpreted as a probability-based sampling rate. Population-expanded analyses are supported through the `trip_weight` field for eligible records.

**Table 3.** Complete Trip Dataset Summary.

| Attribute | Value |
|---|---|
| Study area | Six counties in Utah, USA |
| Study period | January - December 2020 |
| Unique devices | 282,508 |
| Number of trip records | 28,762,367 |
| Number of linked journeys | 28,403,161 |
| Multi-segment reconstructed journeys | 284,336 |
| Multimodal journeys | 218,281 |
| Transit-containing journeys | 994,551 |
| Mean monthly devices | 87,118 |
| Mean monthly trip records | 2,396,825 |
| Travel modes | Car, Bus, Rail, Walk/Bike |
| Spatial representation | Geohash-6 origin and destination cells |
| Temporal representation | 30-minute intervals |
| Network representation | OSM- and GMNS-referenced network paths |
| Population expansion | `trip_weight` for eligible records |
| Number of fields | 20 |
| Linked structure | `trip_id` and `linked_trip_id` |

The released dataset is organized as a trip-level table in which each record corresponds to a single reconstructed trip segment. Complete multimodal journeys are represented by linking multiple trip records through the shared `linked_trip_id`, while individual trip segments are uniquely identified by `trip_id`. **Table 4** summarizes the complete field schema of the released dataset, including trip identifiers, temporal and spatial attributes, travel mode, network representation, transit-specific information, and population expansion weights.

**Table 4.** Field schema of the Complete Trip dataset.

| Field | Description | Example | Note |
|---|---|---|---|
| `linked_trip_id` | Unique identifier for a reconstructed journey containing one or more trip segments. | 202001-9jz6GKQpw4BEy | |
| `trip_id` | Unique identifier for each single-mode trip. | 202002-BwXZAxgpog6 | |
| `segment_order` | Sequential position of the trip segment within the reconstructed journey, starting from 1. | 2 | |
| `travel_mode` | One of four travel modes: 'car', 'walk/bike', 'bus', and 'rail'. | 'car' | |
| `year_start` | Year when trip starts. | 2020 | Specific date is not disclosed |
| `month_start` | Month when trip starts. | 2 | Specific date is not disclosed |
| `day_of_week_start` | Day of week when trip starts. | Mon | Specific date is not disclosed |
| `30m_intrv_start` | 30-minute interval in a day when trip starts. Eligible values are integers from 0 to 47. | 25 | Data type: integer |
| `30m_intrv_end` | 30-minute interval in a day when trip ends. Eligible values are integers from 0 to 47. | 40 | Data type: integer |
| `Geohash6_orig` | 6-digit geohash of trip origin. | 9x0rgp | |
| `Geohash6_dest` | 6-digit geohash of trip destination. | 9x0rgp | |
| `route_taken` | Route taken for the trip. Represented as an ordered list of OSM node IDs (car, walk/bike) or GMNS node IDs (transit). | 84719525, 84846783, 84748783 | |
| `network_distance` | Total reconstructed network distance (miles). | 15.7 | Data type: float |
| `trip_duration` | Reconstructed trip duration in minutes. | 35.5 | Data type: float |
| `route_speed` | Average inferred travel speed (mph). | 32.4, 33.5, 31.2 | |
| `access_stop` | Name of the boarding/access stop in the GMNS transit network. | Sandy Expo | Transit records only |
| `access_stop_id` | `'node_id'` of the access stop. | 1000014 | Transit records only |
| `egress_stop` | Name of the alighting/egress stop in the GMNS transit network. | University Ave / 300 S (NB) | Transit records only |
| `egress_stop_id` | `'node_id'` of the egress stop. | 1000014 | Transit records only |
| `trip_weight` | Population expansion weight for eligible records. | 15.467 | Data type: float |

**Figure 2** illustrates an example complete journey contained in the released dataset. The journey consists of multiple single-mode trip segments connected through a shared `linked_trip_id`, combining rail and bus travel into a single linked journey. This example demonstrates how the dataset preserves complete travel sequences while retaining detailed information for each constituent trip segment, including travel mode, reconstructed network route, transit access and egress stops, and temporal information. The linked-trip structure enables analyses of multimodal travel behavior and complete travel chains while preserving individual trip segments.

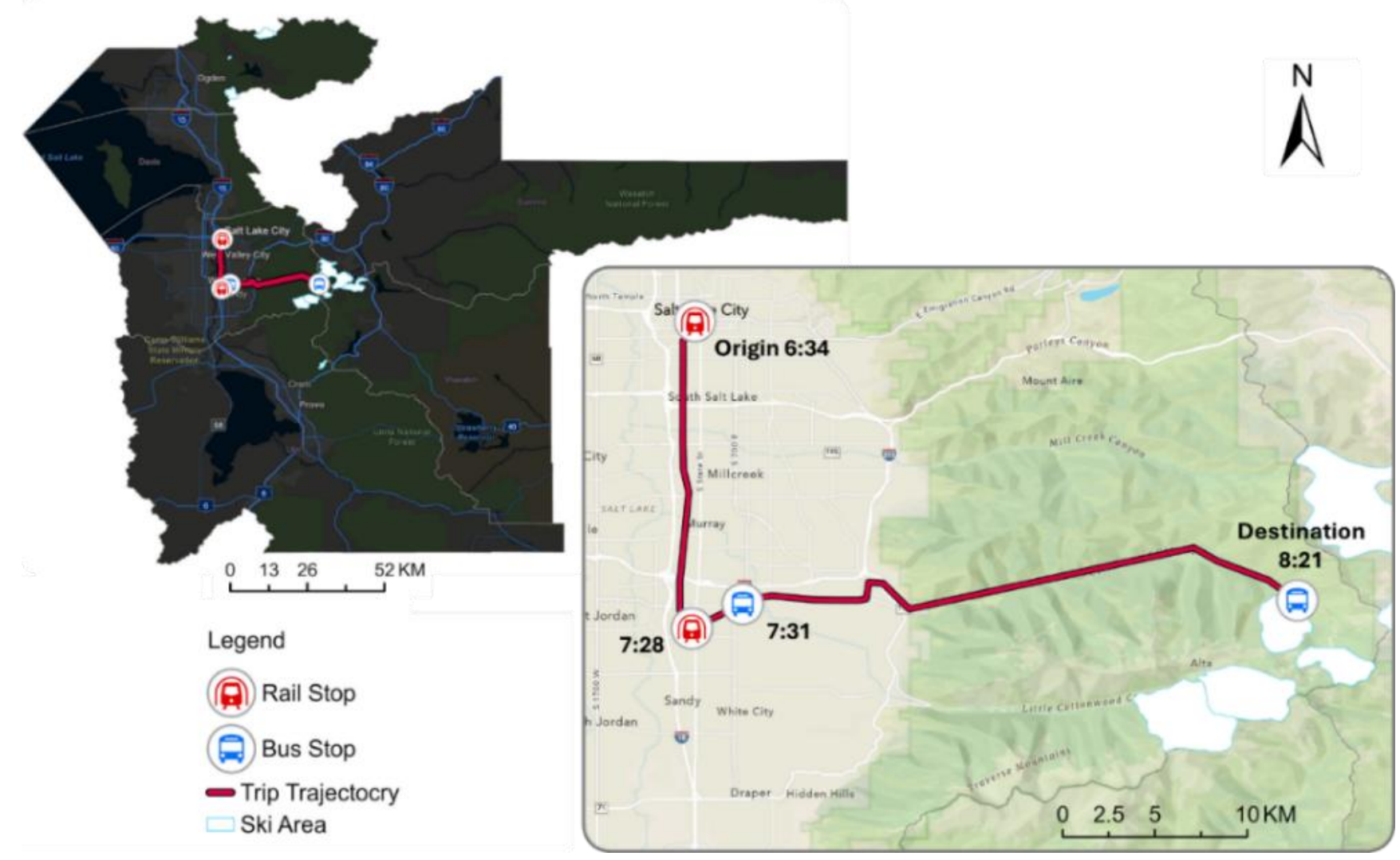

**Figure 2** Example complete journey contained in the released Complete Trip dataset.

**Figure 3** summarizes the descriptive statistics of the released Complete Trip dataset, including travel mode composition, monthly reconstructed trip volumes, and the distributions of trip distance and trip duration by travel mode. These descriptive statistics characterize the overall contents of the released dataset. Quantitative assessments of data quality and reliability are presented in the following Technical Validation section.

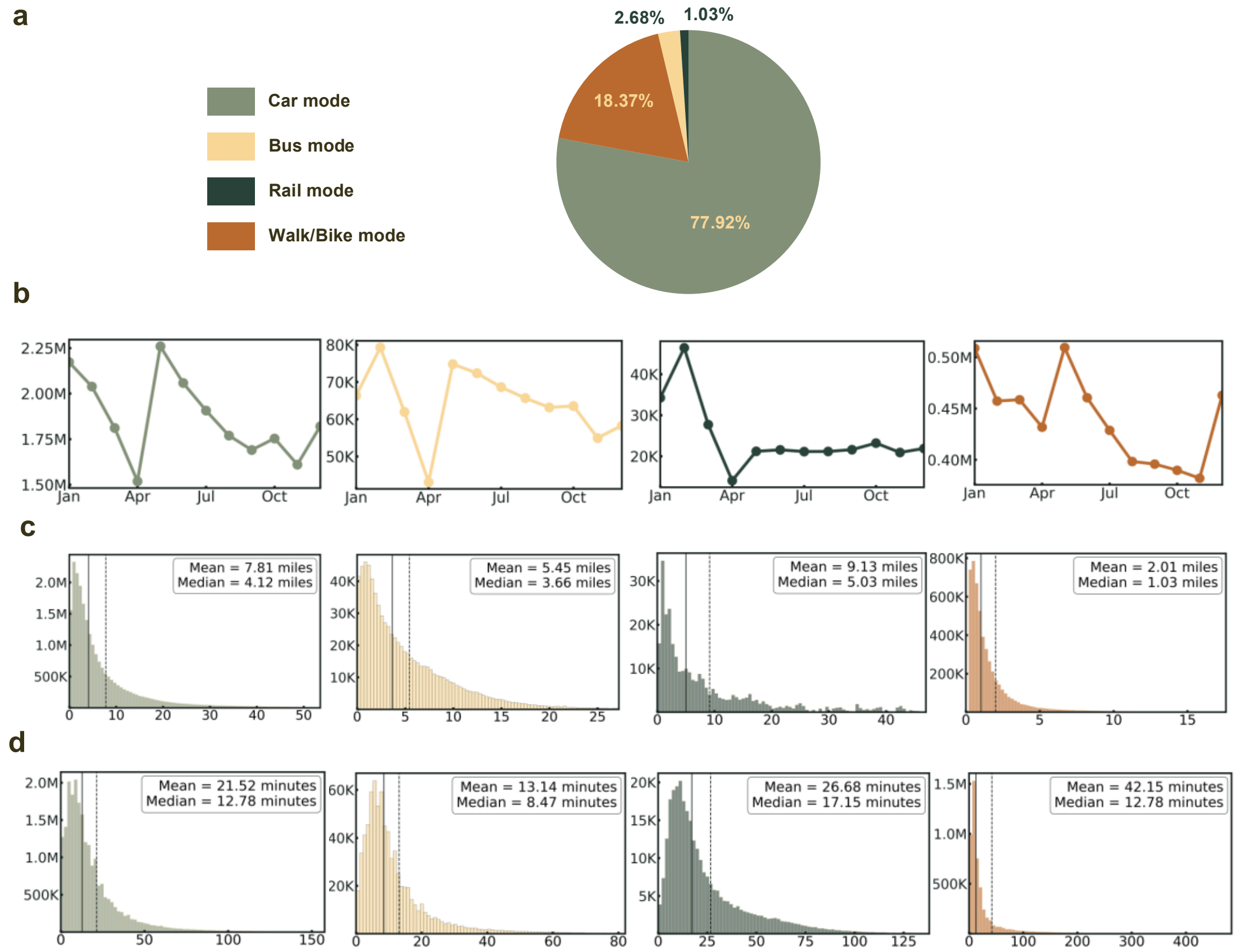

**Figure 3** Descriptive statistics of the released Complete Trip dataset, including (a) travel mode composition, (b) monthly reconstructed trip volumes, (c) trip distance distributions, and (d) trip duration distributions by travel mode.

Together, the dataset summary, field schema, representative linked journey, and descriptive statistics provide a comprehensive overview of the released Complete Trip dataset before its quantitative validation and downstream applications.

## Technical Validation

The reliability of the released Complete Trip dataset was evaluated through three complementary procedures: (1) mode-specific sampling coverage, (2) temporal comparison with official traffic and transit observations, and (3) examination of the completeness and distribution of population expansion weights.

### Sampling rate analysis

**Figure 4** (a), (c), and (e) summarize the estimated sampling rates of car, bus, and rail trips within the Utah six-county study area. Sampling coverage was estimated by comparing reconstructed trips with independent transportation observations, including UDOT traffic counts and UTA transit ridership records. The average sampling rates were 1.32% for car trips, 4.49% for bus trips, and 2.11% for rail trips. Although sampling coverage varies across transportation modes and locations, the observed rates are consistent with those reported for comparable large-scale passive LBS datasets and provide sufficient spatial coverage for corridor- and zone-level mobility analyses.

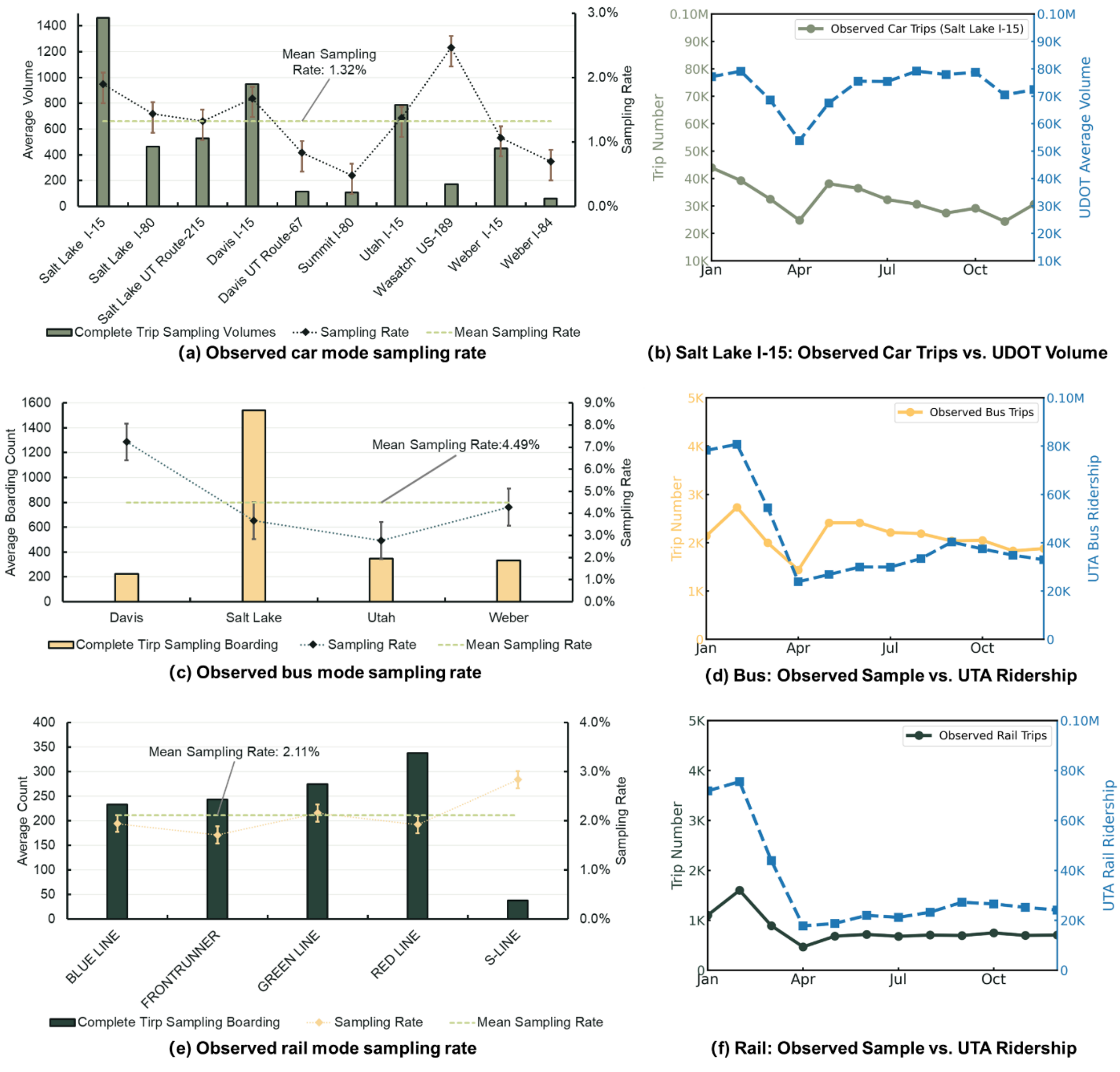

**Figure 4** Technical validation of the released Complete Trip dataset using independent traffic counts and transit ridership observations. (a), (c), and (e) show car, bus, and rail sampling rates in January 2020; (b), (d), and (f) compare monthly trends from the Complete Trip dataset against UDOT and UTA reports.

### Temporal validation

**Figure 4** (b), (d), and (f) compare monthly reconstructed trip volumes with official transportation statistics throughout 2020. Car travel volumes closely follow monthly UDOT traffic counts, reproducing the major temporal variations observed during the study period. Transit validation compares both the unweighted LBS observations and the population-weighted Complete Trip dataset against monthly UTA ridership records. After applying the population expansion weights, the weighted transit trips exhibit a Pearson correlation of $\rho$ = 0.96 with official ridership for both bus and rail modes, improving agreement relative to the unweighted series, as shown in **Figure 4**. These comparisons indicate that the dataset reproduces the principal monthly variations observed in the external benchmarks and that population weighting improves agreement with official transit ridership. Residual sampling and population-coverage biases may nevertheless remain.

### Population weight validation

Valid positive expansion weights were available for 83.90% of trip records. Among records with non-null weights, no non-positive or non-finite values were detected. The median weight was 20.96, and the largest 1% of weights contributed 3.11% of the total weighted mass, indicating limited concentration among extreme weights. The record-level effective sample size was approximately 19.89 million. Records without weights represent categories not eligible for population-level expansion and should not be treated as zero-weight observations.

**Table 5**. Population-weight completeness and distribution

| Metric | Result |
|---|---|
| Trip records with non-null weight | 83.90% |
| Invalid or non-positive weights among non-null weights | 0.00% |
| Mean trip weight | 22.39 |
| Median trip weight | 20.96 |
| 25th percentile | 17.75 |
| 75th percentile | 24.82 |
| 95th percentile | 39.17 |
| 99th percentile | 50.73 |
| Contribution of largest 1% of weights | 3.11% |
| Record-level effective sample size | 19,885,382 |

## Example Applications

The following examples demonstrate how the Complete Trip dataset can support both direct mobility-data products and more advanced transportation research applications. The first group illustrates analyses that can be generated directly from the released trip-level attributes, while the second group presents research-oriented case studies that combine linked journeys, network representations, temporal information, and population expansion weights.

### Illustrative Data Products

**M**ultimodal origin-destination analysis. Figure 5 demonstrates how the released dataset can be used to generate mode-specific OD matrices at the census tract level. Because each reconstructed trip retains travel mode together with privacy-preserving origin and destination locations, users can analyse spatial interaction patterns separately for driving, transit, and active transportation while maintaining a consistent geographic framework across modes. This capability supports multimodal accessibility assessment, travel demand modeling, and comparative analyses of regional mobility patterns.

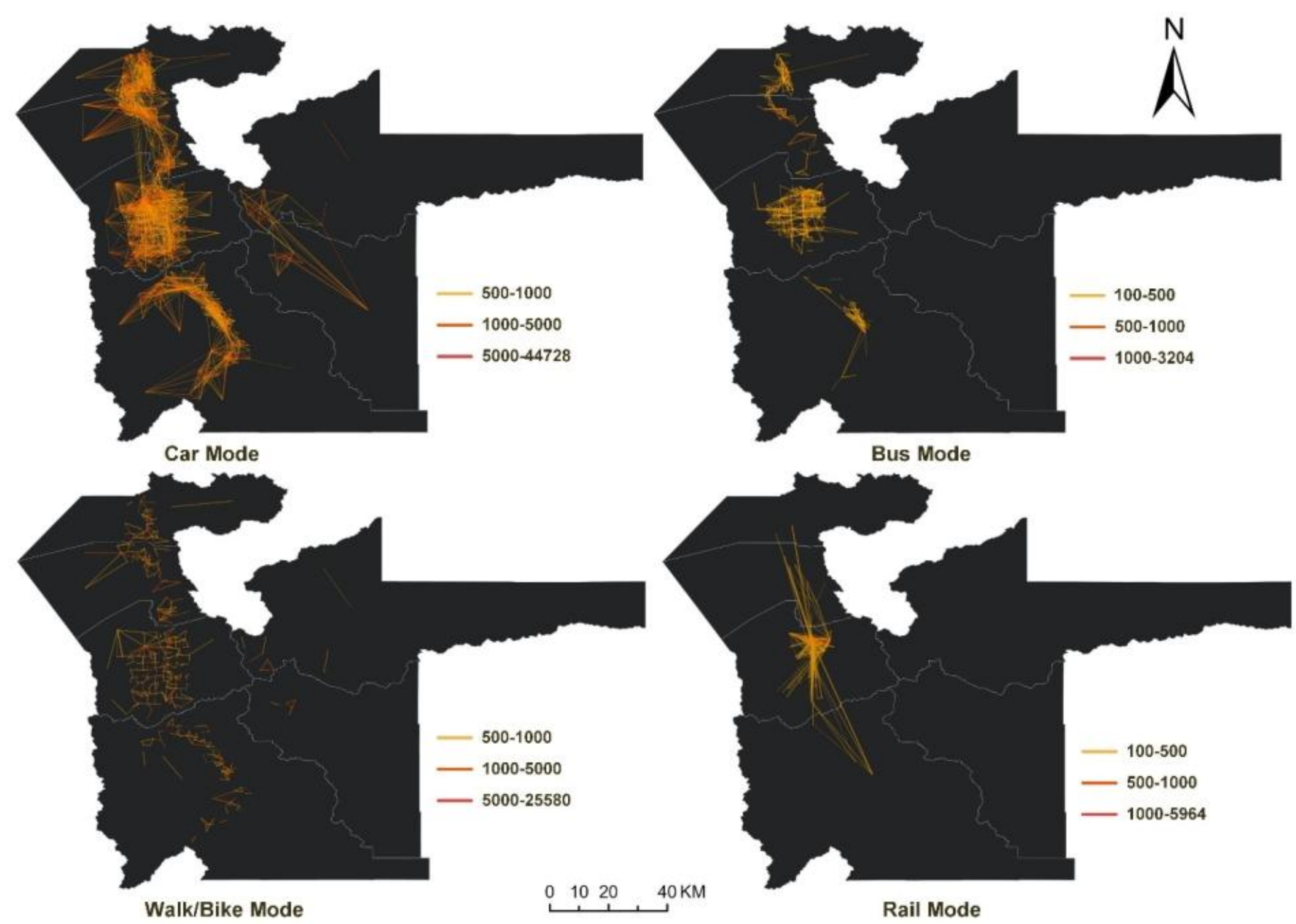

**Figure 5** Mode-Specific Origin-Destination Patterns Across Six Utah Counties in 2020.

**Temporal travel demand analysis. Figure 6** illustrates the use of the released temporal attributes to characterize travel demand throughout the day for weekdays and weekends. The 30-minute temporal resolution enables users to examine diurnal travel patterns, compare temporal demand across transportation modes, and identify peak travel periods while preserving user privacy through temporal aggregation. These data support applications such as transit scheduling, peak-period analysis, and temporal demand forecasting.

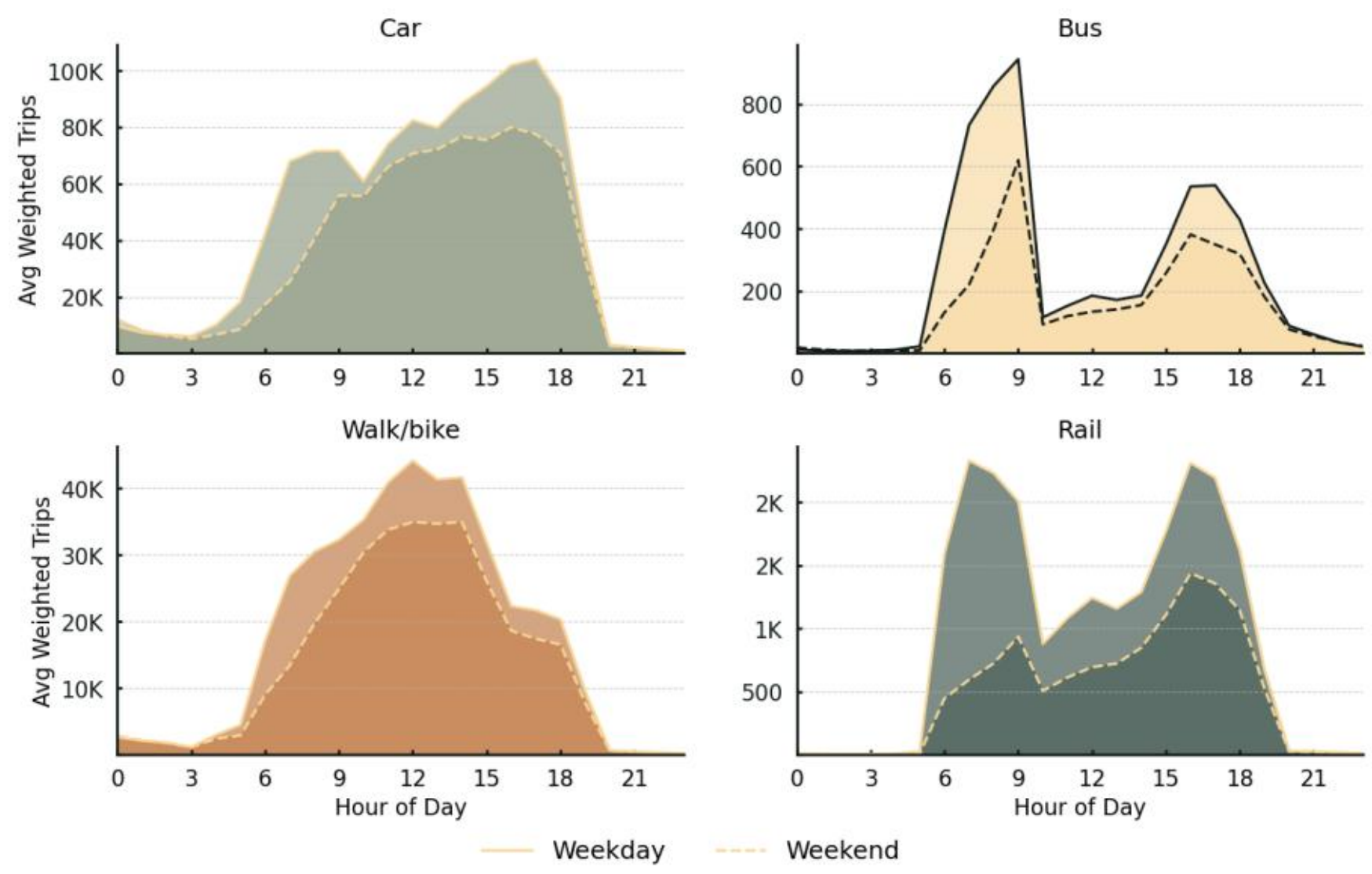

**Figure 6** Hourly Distribution of Trips by Mode on Weekdays and Weekends.

**Network-level traffic analysis. Figure 7** illustrates how reconstructed network routes can be aggregated to estimate traffic volumes on individual road segments. Because each driving and active transportation trip is represented by a map-matched network path, the dataset supports link-level analyses that are not possible using conventional origin-destination matrices alone. Potential applications include traffic demand estimation, corridor performance evaluation, transportation network analysis, and validation of network assignment models.

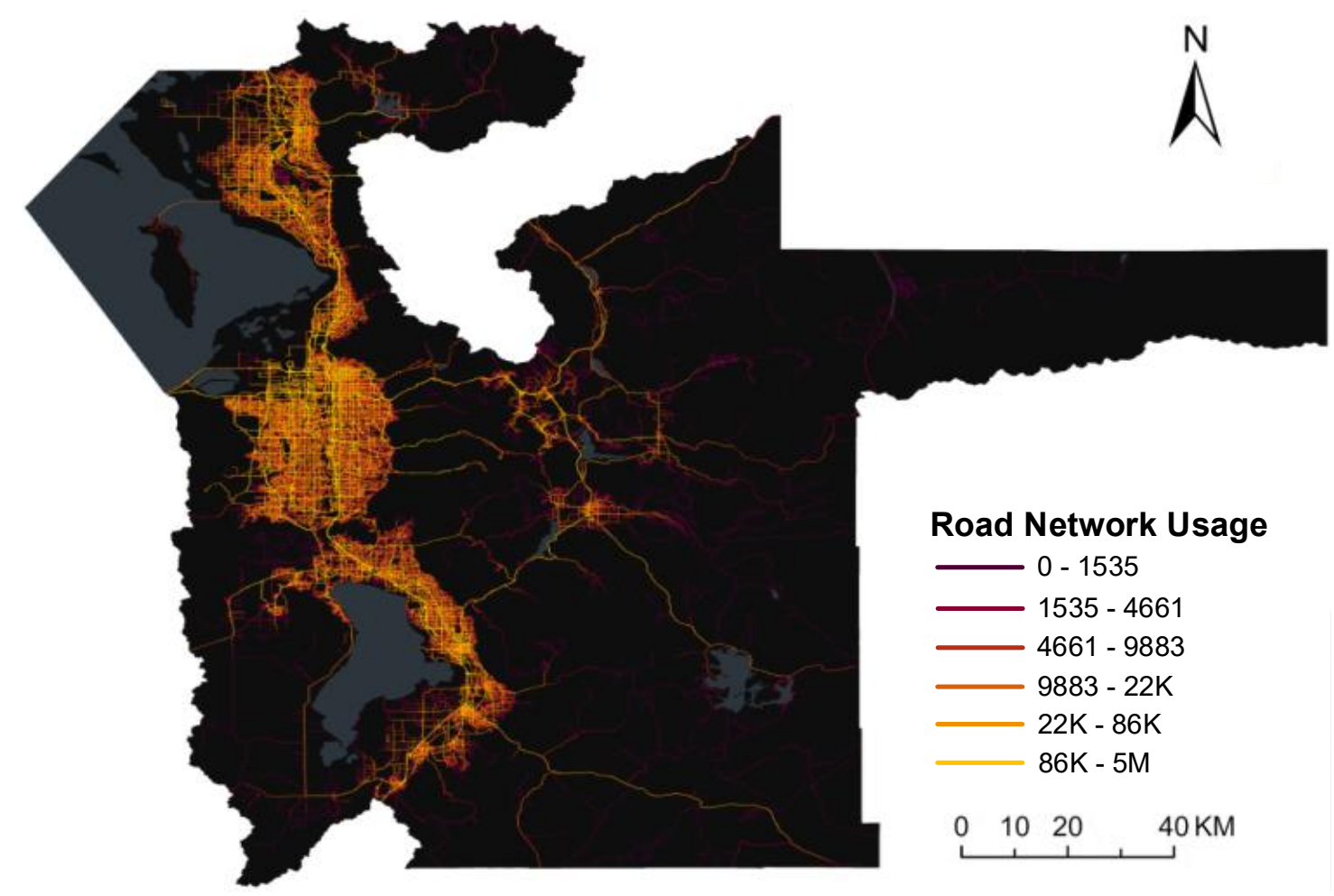

**Figure 7** Estimated Annual Auto Traffic Volume (Unit: Vehicles/Year) 2020.

## Research Application Case Studies

### *Transit desert and mirage analysis*

The Complete Trip dataset supports behaviorally grounded transit-equity analysis by combining population-weighted transit demand with first-mile access information derived from reconstructed journeys (Li et al., 2025). **Figure 8** compares two Transit Mirage examples, the downtown core and an industrial area, with a peripheral residential Transit Desert using first-mile distances to transit access stops. The core and industrial areas exhibit relatively short access distances despite high demand-supply mismatch, indicating that their deficits are not primarily explained by poor physical access to transit. In contrast, the residential desert shows substantially longer first-mile distances, suggesting a more structural accessibility constraint. This case demonstrates how Complete Trip can distinguish transit mismatch associated with travel demand and service use from transit underservice associated with limited first-mile access.

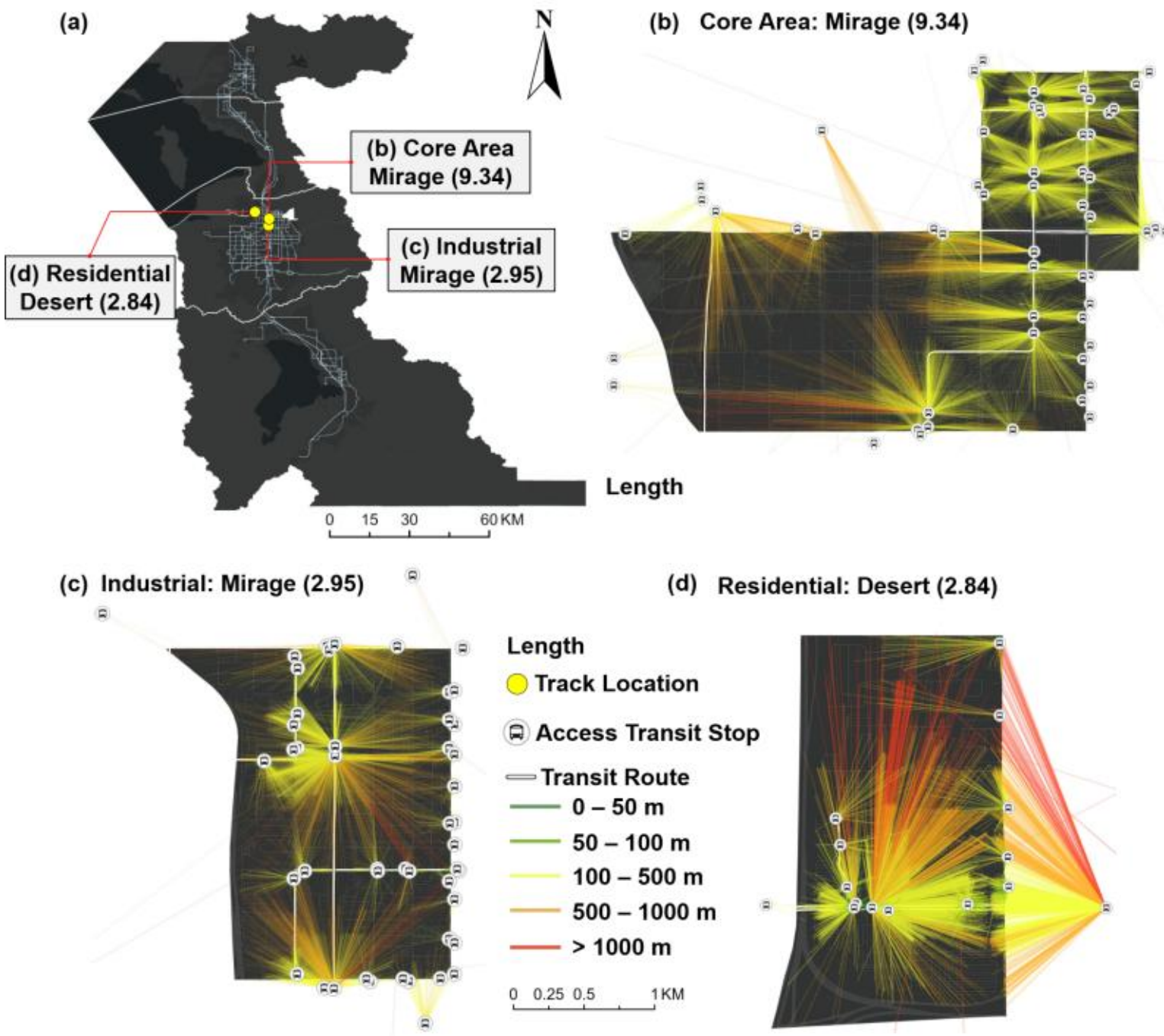

**Figure 8.** Transit Desert and Transit Mirage analysis using Complete Trip.

### *Tourism mobility and visitor travel analysis*

The Complete Trip dataset supports tourism analysis by preserving complete-trip chains before, during, and after visits to major destinations. As illustrated in **Figure 9**, ski-resort visitors were classified as hotel guests, Airbnb guests, or residents using inferred home locations, and their resort visits were linked with subsequent dining, shopping, recreation, and entertainment trips. The reconstructed activity sequences demonstrate how Complete Trip can link destination visits with subsequent activities and compare daily travel patterns across visitor groups. This application highlights the dataset's ability to support visitor classification, activity-sequence reconstruction, travel-time analysis, destination-based trip chaining, and cross-destination mobility analysis beyond aggregate visit counts.

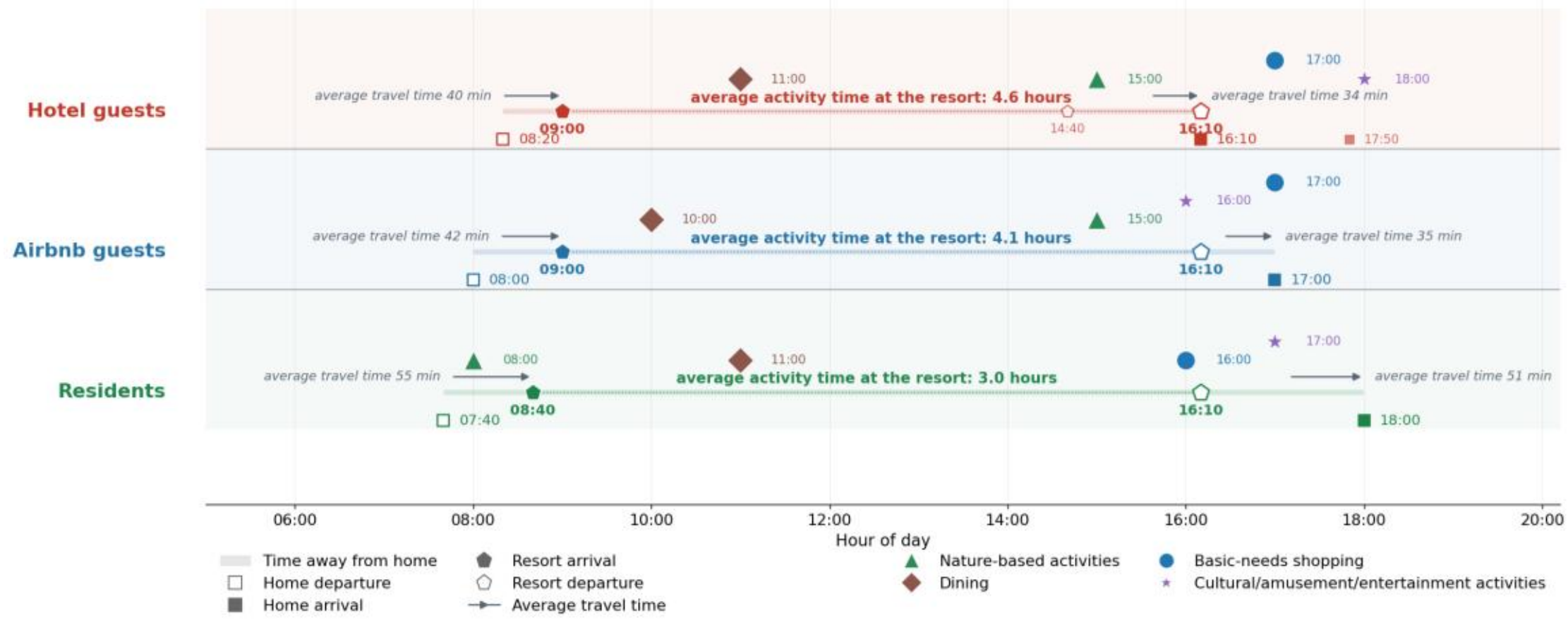

**Figure 9.** Tourism daily activity sequences reconstructed from Complete Trip.

## Usage Notes

The Complete Trip dataset is intended primarily for population-level analyses of multimodal human mobility. The following notes highlight key fields and recommended practices for appropriate interpretation and reuse of the released dataset.

**Population expansion weights.** Analyses intended to represent the traveling population should apply the `trip_weight` field, which incorporates both device-level expansion and trip-level calibration factors. Weighted analyses provide population-expanded estimates calibrated to the selected benchmarks, although residual biases may remain. Records with null `trip_weight` values should be excluded from weighted aggregation rather than assigned a value of zero.

**Origin and destination coordinates.** The `Geohash6_orig` and `Geohash6_dest` fields report trip origins and destinations as Geohash-6 cell identifiers rather than precise coordinates. These generalized locations should not be interpreted as exact departure addresses, building-level locations, or GPS-recorded coordinates. Spatial analyses should be conducted at the Geohash-6 resolution or at coarser aggregation levels, such as census tracts.

**Temporal fields.** Trip start and end times are reported as 30-minute interval indices within a day rather than exact timestamps. For example, a `30m_intrv_start` value of 16 corresponds to the 8:00-8:30 AM interval. Exact calendar dates are not released; temporal information is represented by `year_start`, `month_start`, `day_of_week_start`, `30m_intrv_start`, and `30m_intrv_end`. Analyses requiring exact travel times or precise date-level sequencing cannot be supported by the released data.

**Route representation.** The `route_taken` field encodes each trip's map-matched path as an ordered sequence of OSM node IDs (for car, walk, and bike trips) or GMNS node IDs (for transit trips). The reconstructed path should be interpreted as an inferred network route rather than the device's original GPS trajectory. To map node ID sequences back to geographic geometries, users should reference the accompanying OSM-derived road network files or the GTFS-derived transit network, both available in the dataset repository.

**Linked trip chains.** Individual trip segments belonging to the same complete journey share a common `linked_trip_id`. Users can reconstruct journey groups by grouping records with the same `linked_trip_id` and ordering them by `segment_order` to reconstruct the complete journey sequence. Transit-specific fields (`access_stop`, `access_stop_id`, `egress_stop`, and `egress_stop_id`) are populated only for transit segments within linked journeys, whereas non-transit segments contain null values for these fields.

## Code and Data Availability

No custom software is required to use the released dataset. The mobility reconstruction framework follows the standardized methodology developed for NextGen NHTS (FHWA, 2021). The accompanying data repository provides the dataset documentation, field definitions, and supporting network files required for data access and analysis.

The Complete Trip dataset is released as a privacy-preserving, population-weighted multimodal mobility dataset for the Utah six-county region (Salt Lake, Davis, Weber, Utah, Wasatch, and Summit counties) covering calendar year 2020. A representative sample dataset, together with the complete field schema, documentation, supporting OSM/GMNS network files, and usage notes, is publicly available through the project GitHub repository: https://github.com/villanova-transportation/Complete-Trip-Data. An interactive online data explorer is also provided at https://villanova-transportation.github.io/Complete-Trip-Data-Explorer/explorer.html.
Supporting road, pedestrian, and transit network files are provided in GMNS format to facilitate interpretation of the released `route_taken` field and network-based analyses.

Access to the complete Utah dataset is available to qualified researchers upon request through the NovaMobility Lab at Villanova University under a data use agreement. Data access requests should be submitted by email to Chenfeng Xiong at chenfeng.xiong@villanova.edu.

The proposed data production framework is geographically transferable and can be applied to other metropolitan or regional study areas with sufficient LBS coverage. Researchers interested in producing Complete Trip datasets for additional regions are encouraged to contact the corresponding authors.

## Author contributions

W.L. developed the dataset generation framework, implemented the mobility reconstruction pipeline, and contributed to manuscript writing.

R.L. contributed to methodology development with a focus on route reconstruction and map matching, led the manuscript development and dataset documentation, and conducted data analysis and technical validation.

X.W. contributed to methodology development and supported analytical refinement.

C.X. conceptualized the study, secured resources, and supervised the project.

All authors reviewed and approved the final manuscript.

## References

1. De Montjoye, Y.A., Hidalgo, C.A., Verleysen, M. and Blondel, V.D., 2013. Unique in the crowd: The privacy bounds of human mobility. *Scientific reports*, *3*(1), p.1376.
2. Federal Highway Administration (FHWA), 2022. Complete Trip Data Factsheet. [pdf] Washington, D.C.: U.S. Department of Transportation. Available at: https://ops.fhwa.dot.gov/publications/fhwahop22066/fhwahop22066.pdf

3. Federal Highway Administration (FHWA) (2025) *Office of Highway Policy Information - Policy | Travel Monitoring (TVT)*. Available at: https://www.fhwa.dot.gov/policyinformation/travel_monitoring/tvt.cfm (Accessed: 30 July 2025).
4. Federal Highway Administration, (2021). Next generation National Household Travel Survey National Origin Destination Data Passenger Origin-Destination Data Methodology Documentation. *nhts.ornl.gov/od/assets/doc/2020_NextGen_NHTS_Passenger_OD_Data_Methodology_v2.pdf*. Accessed 01, Jan 2025.
5. Google LLC, 2020. COVID-19 community mobility reports. URL: https://www.google.com/covid19/mobility/.
6. Gonzalez, M.C., Hidalgo, C.A. and Barabasi, A.L., 2008. Understanding individual human mobility patterns. *Nature*, *453*(7196), pp.779-782.
7. Hadi, M., Zhou, X. and Hale, D., 2022. *Multiresolution modeling for traffic analysis:* Guidebook (No. FHWA-HRT-22-055). United States. Federal Highway Administration.
8. Hu, S., Xiong, C., Yang, M., Younes, H., Luo, W. and Zhang, L., 2021. A big-data driven approach to analyzing and modeling human mobility trend under non-pharmaceutical interventions during COVID-19 pandemic. *Transportation Research Part C: Emerging Technologies*, *124*, p.102955.
9. Kashiyama, T., Pang, Y. and Sekimoto, Y., 2017. Open PFLOW: Creation and evaluation of an open dataset for typical people mass movement in urban areas Transportation Research Part C: Emerging Technologies, 85, pp.249-267.
10. Li, R., Xiong, C., Tavakoli, A., Nataraj, C, 2024. Map Matching of Location Data Trajectories: A Heterogeneous and Bayesian-Optimized Hidden Markov Approach. Presented at Conference in Emerging Technologies in Transportation Systems (TRC-30), Crete, Greece.
11. Li, R., Luo, W. & Xiong, C., 2025. Equity in motion: Rethinking transit deserts through complete trip data. Presented at the AGU 2025 Annual Meeting, New Orleans, Louisiana, 15-19 December 2025.
12. Liu, K., Wu, X., Molnar, L., Hu, S., Zhang, L. and Xiong, C., 2025. Characterizing human mobility and its impact on the spread of COVID-19 during extreme weather events: A case study of hurricane Laura. *Journal of Transport Geography*, *129*, p.104423.
13. Molloy, J., Castro, A., Götschi, T., Schoeman, B., Tchervenkov, C., Tomic, U., Hintermann, B., Axhausen, K.W., 2023. The MOBIS dataset: A large GPS dataset of mobility behaviour in Switzerland. Transportation 50, 1983–2007. https://doi.org/10.1007/s11116-022-10299-4.
14. Monreale, A. and Pellungrini, R., 2023. A survey on privacy in human mobility. *Transactions on Data Privacy*, *16*(1), pp.51-82.
15. National Household Travel Survey (2025) *National Household Travel Survey*. Available at: https://nhts.ornl.gov/ (Accessed: 30 May 2025).
16. New York City Taxi and Limousine Commission (TLC), 2025. *TLC Trip Record Data*. [online] Available at: https://www.nyc.gov/site/tlc/about/tlc-trip-record-data.page [Accessed 29 July 2025].
17. OpenStreetMap (2024) *OpenStreetMap*. Available at: https://www.openstreetmap.org/ (Accessed: 26 January 2024).
18. SafeGraph, 2022. Determining points of interest visits from location data: A technical guide to visit attribution. SafeGraph Resources. URL: https://www.safegraph.com/guides/visit-attribution-white-paper.
19. Song, C., Qu, Z., Blumm, N. and Barabási, A.L., 2010. Limits of predictability in human mobility. *Science*, *327*(5968), pp.1018-1021.
20. Sun, J., Guo, J., Wu, X., Zhu, Q., Wu, D., Xian, K. and Zhou, X., 2019. Analyzing the impact of traffic congestion mitigation: From an explainable neural network learning framework to marginal effect analyses. Sensors, 19(10), p.2254.

21. Sinclair, C. and Schaefer, R., 2019. Multimodal System Performance Measures Research and Application: Innovation and Research Plan (No. FHWA-HOP-18-085). United States. Federal Highway Administration.
22. U.S. Department of Transportation (2025) *Complete Monthly Ridership (with Adjustments and Estimates)*. Available at: https://data.transportation.gov/Public-Transit/Complete-Monthly-Ridership-with-Adjustments-and-Es/8bui-9xvu/about_data.
23. Wang, F., Chen, C., 2018. On data processing required to derive mobility patterns from passively-generated mobile phone data. Transportation Research Part C: Emerging Technologies 87, 58–74. https://doi.org/10.1016/j.trc.2017.12.003.
24. Wang, F., Wang, J., Cao, J., Chen, C., Ban, X.J., 2019. Extracting trips from multi-sourced data for mobility pattern analysis: An app-based data example. Transportation Research Part C: Emerging Technologies 105, 183–202. https://doi.org/10.1016/j.trc.2019.05.028.
25. Wu, X., Guo, J., Xian, K. and Zhou, X., 2018. Hierarchical travel demand estimation using multiple data sources: A forward and backward propagation algorithmic framework on a layered computational graph. Transportation Research Part C: Emerging Technologies, 96, pp.321-346.
26. Xiong, C., Yang, D. and Zhang, L., 2018. A high-order hidden Markov model and its applications for dynamic car ownership analysis. *Transportation Science*, *52*(6), pp.1365-1375.
27. Xiong, C., Hu, S., Yang, M., Luo, W. and Zhang, L., 2020. Mobile device data reveal the dynamics in a positive relationship between human mobility and COVID-19 infections. *Proceedings of the National Academy of Sciences*, *117*(44), pp.27087-27089.
28. Yabe, T., Tsubouchi, K., Shimizu, T., Sekimoto, Y., Sezaki, K., Moro, E. and Pentland, A., 2024. YJMob100K: City-scale and longitudinal dataset of anonymized human mobility trajectories. Scientific Data, 11(1), p.397.
29. Yang, M., Pan, Y., Darzi, A., Ghader, S., Xiong, C., Zhang, L., 2022. A data-driven travel mode share estimation framework based on mobile device location data. Transportation 49, 1339–1383. https://doi.org/10.1007/s11116-021-10214-3.
30. Yin, L., Wang, Q., Shaw, S.L., Fang, Z., Hu, J., Tao, Y. and Wang, W., 2015. Re-identification risk versus data utility for aggregated mobility research using mobile phone location data. *PloS One*, *10*(10), p.e0140589.
31. Zhou, X., Hadi, M. and Hale, D.K., 2021. Multiresolution Modeling for Traffic Analysis: State-of-Practice and Gap Analysis Report (No. FHWA-HRT-21-082). United States. Federal Highway Administration.
32. Zhou, X.S., Cheng, Q., Wu, X., Li, P., Belezamo, B., Lu, J. and Abbasi, M., 2022. A meso-to-macro cross-resolution performance approach for connecting polynomial arrival queue model to volume-delay function with inflow demand-to-capacity ratio. Multimodal Transportation, 1(2), p.100017.
33. Zheng, Y., Wang, L., Zhang, R., Xie, X. and Ma, W.Y., 2008, April. GeoLife: Managing and understanding your past life over maps. In The Ninth International Conference on Mobile Data Management (mdm 2008) (pp. 211-212). IEEE.
34. Zhu, Y., Ye, Y., Wu, Y., Zhao, X. and Yu, J., 2023. Synmob: Creating high-fidelity synthetic GPS trajectory dataset for urban mobility analysis. *Advances in Neural Information Processing Systems*, *36*, pp.22961-22977.
35. Zephyr Foundation (n.d.) *General Modeling Network Specification (GMNS)*. Available at: https://github.com/zephyr-data-specs/GMNS (Accessed: 1 June 2025).
36. Zhang, L., Wu, X.B., Liu, K., Al Mehedi, M.A., Zhou, J., Smith, V. and Xiong, C., 2026. Diagnostic framework for causal inference in seasonal urban flooding: Precipitation-based control selection and synthetic difference-in-differences in Lagos, Nigeria. *International Journal of Disaster Risk Reduction*, p.105993.